\documentclass[]{spie}  %>>> use for US letter paper
\usepackage{amsmath,amsfonts,amssymb}
\usepackage{graphicx}
\usepackage[colorlinks=true, allcolors=blue]{hyperref}

\title{Development of an integrated near-IR astrophotonic spectrograph}

\author[a,b]{Pradip Gatkine}
\author[b,c]{Meghna Sitaram}
\author[b,d]{Sylvain Veilleux}
\author[e]{Mario Dagenais}
\author[f]{Joss Bland-Hawthorn}

\affil[a]{Division of Physics, Mathematics and Astronomy, California Institute of Technology, Pasadena, CA 91125, USA}
\affil[b]{Department of Astronomy, University of Maryland, College Park, Maryland 20742, USA}
\affil[c]{Department of Astronomy, Columbia University, New York, New York 10027, USA}
\affil[d]{Joint Space-Science Institute, University of Maryland, College Park, Maryland 20742, USA}
\affil[e]{Department of Electrical and Computer Engineering, University of Maryland, College Park, Maryland 20742, USA }
\affil[f]{Sydney Institute for Astronomy and Sydney Astrophotonic Instrumentation Labs, School of Physics, The University of Sydney, New South Wales 2006, Australia}

\authorinfo{Further author information: (Send correspondence to P. Gatkine)\\E-mail: pgatkine@caltech.edu}

\begin{document} 
\maketitle

\begin{abstract}
Here, we present an astrophotonic spectrograph in the near-IR H-band (1.45 -1.65 $\mu m$) and a spectral resolution ($\lambda/\delta\lambda$) of 1500. The main dispersing element of the spectrograph is a photonic chip based on Arrayed-Waveguide-Grating technology. The 1D spectrum produced on the focal plane of the AWG contains overlapping spectral orders, each spanning a 10 nm band in wavelength. These spectral orders are cross-dispersed in the perpendicular direction using a cross-dispersion setup which consists of collimating lenses and a prism and the 2D spectrum is thus imaged onto a near-IR detector. Here, as a proof of concept, we use a few-mode photonic lantern to capture the light and feed the emanating single-mode outputs to the AWG chip for dispersion. The total size of the setup is 50$\times$30$\times$20 cm$^3$, nearly the size of a shoebox. This spectrograph will pave the way for future miniaturized integrated photonic spectrographs on large telescopes, particularly for building future photonic multi-object spectrographs. 
\end{abstract}

% Include a list of keywords after the abstract 

%\section{INTRODUCTION}
%\label{sec:intro}  % \label{} allows reference to this section
\section{Introduction}
The photonic platform of guided light in fibers and waveguides has opened the doors to next-generation instrumentation for both ground- and space-based telescopes in optical and near/mid-IR bands, particularly for the upcoming extremely large telescopes (ELTs) \cite{bland2009astrophotonics, bland2017astrophotonics, ellis2017astrophotonics, gatkine2019astrophotonic, Gatkine2019State}. The key idea here is to leverage the ability of guiding the light in waveguides (using total internal reflection) to collapse the conventional optical setups into 2D ``optical circuits". These optical circuits make use of single-mode waveguides or fibers, which are equivalent to diffraction-limited operation. The single-mode operation allows highly flexible and unique manipulations of light (such as fine/active control of the path lengths) which are not feasible with conventional optics. Thanks to the 2D nature and single-mode propagation, the photonic chips/devices are highly miniaturized (particularly spectrographs). 
The size of these miniaturized photonic chips remains the same irrespective of the diameter of the telescope. Therefore, they can break the spiral of cost, weight, and volume (which scale as $D^{2+}$), which poses a major challenge for spectrograph/instrument design and its thermal/mechanical stability as the telescope diameter ($D$) grows \cite{bland2009astrophotonics}. This makes the photonic devices and spectrographs particularly useful for the upcoming ELTs (eg: the Thirty Meter Telescope) as well as space-based and balloon-based telescopes\cite{bland2017astrophotonics, blind2017spectrographs, bland2006instruments}.

\section{An Integrated Photonic Spectrograph}

An integrated photonic spectrograph consists of a photonic chip as the main disperser, a single or a set of single-mode fibers feeding the photonic chip, and an imaging or a cross-dispersing setup to capture the spectrum on a detector. The on-chip photonic implementations are of interest since they are highly compact (chips are only a few cm$^2$ in size), stable, flexible in terms of specifications, and easily stackable to create an IFU or multi-object spectrograph \cite{cvetojevic2012first, harris2012applications, gatkine2017arrayed}.
Some of the main photonic dispersion techniques include photonic echelle gratings (PEGs), arrayed waveguide gratings (AWGs), and Fourier-transform spectrometer (FTS). These techniques are described in more detail from an astronomical perspective in a recent review paper \cite{gatkine2019astrophotonic}.  Of these techniques, AWG is the most promising for astronomical spectroscopy, thanks to its ease of fabrication, flexibility, and relatively higher throughput that can be achieved.  

An AWG uses an on-chip phased-array-like structure of waveguides to introduce progressive path lengths similar to a grating. These discrete light paths with increasing phase delays then create an interference pattern at the output free propagation region with different wavelengths within a spectral order constructively interfering at different locations on the focal line (or focal curve) \cite{gatkine2017arrayed}.    

%% Adding AWG-conventional-similarity
   \begin{figure} [ht]
   \begin{center}
   \begin{tabular}{c} %% tabular useful for creating an array of images 
   \includegraphics[height=5.5cm]{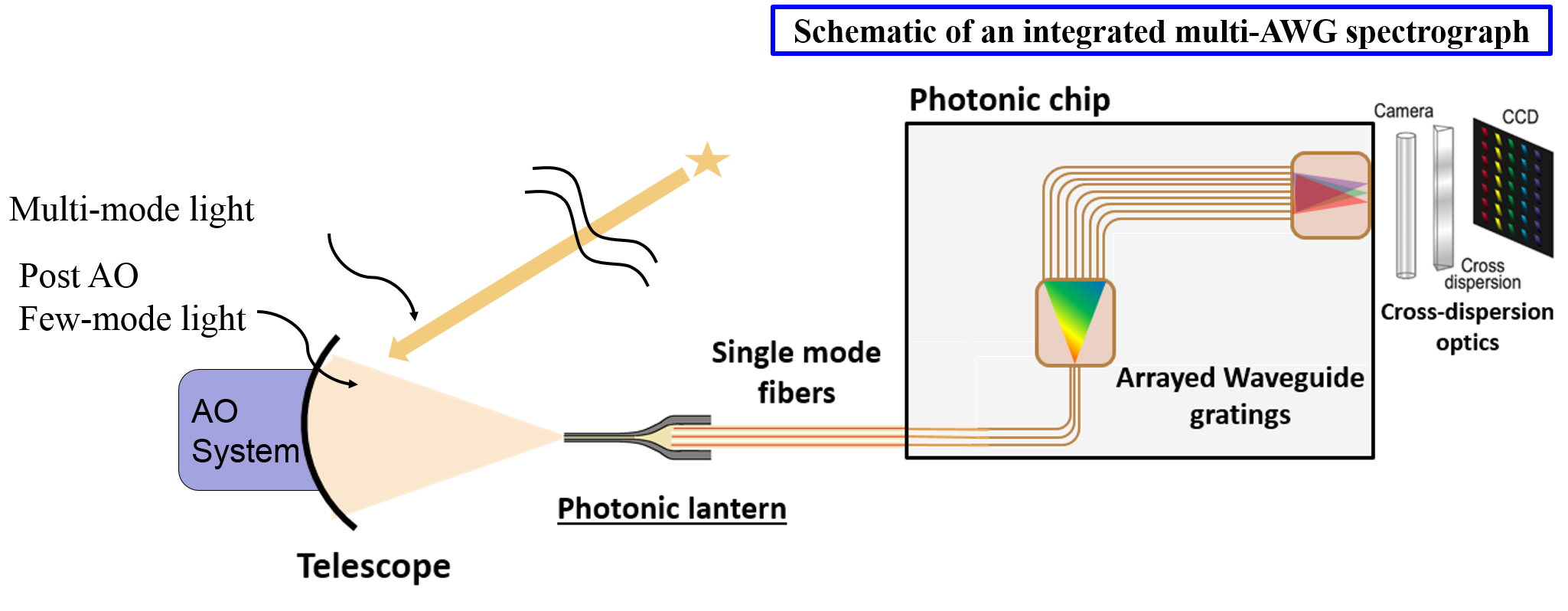}
   \end{tabular}
   \end{center}
   \caption[Full setup of an integrated photonic spectrograph] 
%>>>> use \label inside caption to get Fig. number with \ref{}
   { \label{fig:Full_setup} 
Full setup of an integrated photonic spectrograph with photonic lanterns, multi-input AWG, and cross-dispersion system. Here, the single-mode fibers feed the single-mode waveguides on the chip, which in turn illuminate the input free propagation region (shown as a maroon rectangle on the left). The phase differences are introduced in the discrete array of waveguides, which then interfere to create a spectrum in the output free propagation region (shown as a maroon rectangle on the right). The cross-dispersion setup separates the overlapping spectral orders in a direction perpendicular to the chip and images the spectrum onto a detector. 
The waveguide Bragg Grating filters \cite{zhu2016arbitrary, hu2020integrated} can be introduced in the input waveguides of the AWG for filtering out atmospheric emission lines before they illuminate the free propagation region of the AWG.}
   \end{figure} 
  
Figure \ref{fig:Full_setup} shows the schematic of an integrated photonic spectrograph using the AWG chip as the primary disperser. The light received at the ground-based telescope is multi-moded (i.e. seeing limited). The AO system is used to reduce the degrees of freedom of the focused light to create a focused spot which is closer to the diffraction limit (few-moded light). The light can be directly focused on a single-mode fiber (in case of high Strehl ratio extreme-AO systems, as previously demonstrated \cite{jovanovic2017demonstration}) or can be focused on a few-mode photonic lantern\cite{leon2005multimode, horton2007coupling} which then adiabatically guides the light into several single-mode fibers (as previously demonstrated \cite{ellis2020first}). The AWG chip acts as a primary disperser and disperses the light along the focal line as shown in Fig. \ref{fig:Full_setup}. However, multiple spectral orders overlap at the focal line. The spectral orders are separated perpendicular to the chip using a low-resolution cross-dispersion setup to image a 2-D spectrum onto a detector.

In Gatkine et al. 2017 \cite{gatkine2017arrayed}, we demonstrated a single-input, moderate resolution (R $\sim$ 1500), high-throughput (peak overall throughput $\sim$ 25\%), broad-band AWG designed for H-band (1450$-$1550 nm) and a multi-input AWG was described in Gatkine et al. 2018 \cite{gatkine2018towards}. We also demonstrated efficient waveguide tapers ($>$ 90\% coupling efficiency) to couple the single-mode fiber modes (diameter $\sim$ 10 $\mu m$) with single-mode waveguides on the chip (cross section $\sim$ 2 $\times$ 0.1 $\mu m^{2}$) \cite{zhu2016ultrabroadband, gatkine2016development, hu2018characterization}. For this AWG spectrograph, we focused on astronomical H band (1.45 – 1.65 $\mu m$) for two reasons: 1) The UV/optical light from the stars in the earliest galaxies ($z > 6$) is redshifted to this waveband. 2) This waveband is widely used in telecommunication applications and hence, the material properties are well-known and various peripherals (eg: fibers, lasers, spectrum analyzers) are readily available. In this paper, we discuss the cross-dispersion system and integration of this chip as a 2D sepctrograph.

\section{Cross-dispersion setup}

This AWG chip \cite{gatkine2017arrayed} has a resolving power of ($\lambda/\delta\lambda$ $\sim$ 1500) and a free spectral range of 10 nm. Thus the cross-dispersion setup is used to separate the $\sim$20 overlapping spectral orders (bandwidth/FSR = 200 nm / 10 nm = 20 spectral orders).  
%The cross-dispersion setup is to be used to separate the overlapping spectral orders at the output waveguides of the AWG or its exposed focal plane (as discussed in Section \ref{sec:FPR_exposed}). 
In this section, we summarize the detector characterization and cross-dispersion system, which will then be coupled with the AWG chip described above.

\begin{figure} [tbp]
   \begin{center}
   \begin{tabular}{c} %% tabular useful for creating an array of images 
   \includegraphics[height=7cm]{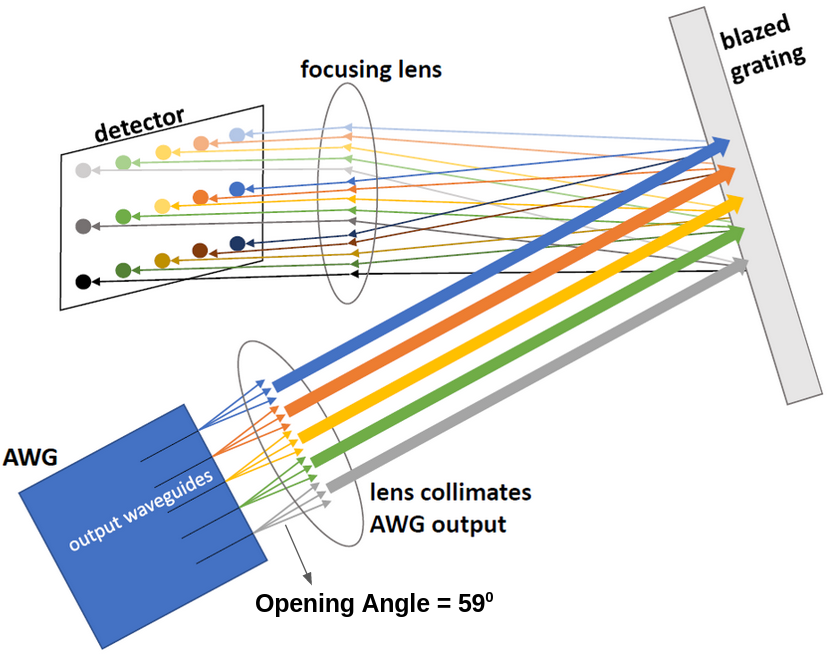}
   \end{tabular}
   \end{center}
   \caption[AWG-cross-dispersion schematic] 
%>>>> use \label inside caption to get Fig. number with \ref{}
   { \label{fig:AWG_to_detector} 
A schematic showing the AWG $+$ cross-dispersion assembly.  The AWG output facet will be mounted at the focal plane of the collimating lens. Each bright color indicates the finer dispersion by the AWG (with $\delta \lambda$ $\sim$ 1 nm), albeit with overlapping spectral orders. The blazed grating performs a coarse dispersion ($\Delta \lambda$ $\sim$ 10 nm) and thus, separates the spectral orders in the orthogonal direction (as shown by the shades of the colors).}
   \end{figure}

\subsection{Detector Characterization}
Considering the separation of 20-23 overlapping spectral orders (each order with a  free spectral range,  $\Delta\lambda$ $\sim$ 10 nm), we assumed roughly 5- to 6-pixel separation between the orders, thus requiring a 2D detector array with more than 128 $\times$ 128 pixels. Our choice of the 2D detector array for the preliminary demonstration of cross-dispersed AWG was mostly driven by the cost. Thus, we chose Hamamatsu G12242-0707W image sensor, an Indium Gallium-Arsenide (InGaAs) 2D detector array with dimensions of 128 x 128 pixels and a pixel pitch of 20 $\mu m$ \footnote{The complete datasheet can be found here:\\ https://www.hamamatsu.com/resources/pdf/ssd/g12242-0707w\_kmir1022e.pdf}. The quantum efficiency of the detector in the H-band is 60\%. We use the detector in conjunction with a  C11512-02  detector head from Hamamatsu which includes the circuitry for driver, controller, thermo-electric cooler, and readout\footnote{The complete datasheet can be found here: \\ https://www.hamamatsu.com/resources/pdf/ssd/c11512\_series\_kacc1194e.pdf}. The data and control instructions are transferred using a high-speed CameraLink interface. The image readout and control is performed using a software called DCAM-CL. Before any image acquisition, we perform a dark calibration by capturing an image with the detector covered.   

\begin{figure} [tbp]
   \begin{center}
   \begin{tabular}{c} %% tabular useful for creating an array of images 
   \includegraphics[height=6cm]{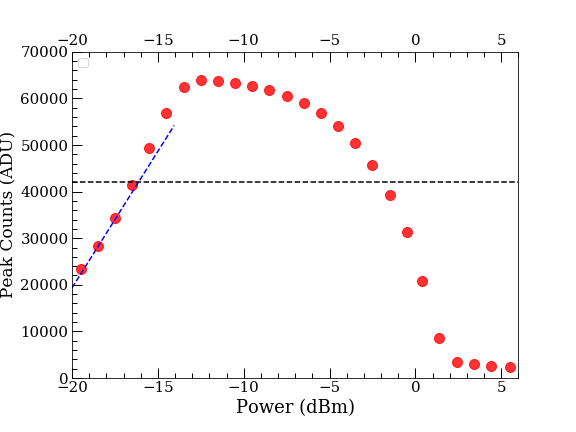}
   \end{tabular}
   \end{center}
   \caption[Saturation characteristics] 
%>>>> use \label inside caption to get Fig. number with \ref{}
   { \label{fig:saturation_characteristics} 
Saturation characteristics of the detector. The detector is in the linear regime upto 42000 ADU (as shown by the dashed lines). The saturation level is 65535 ADU.}
   \end{figure}

We measure the saturation characteristics of the detector to determine the range of linear operation. This test was performed at $\lambda$ = 1450 nm. For this test, a laser source fed a single-mode fiber. The fiber was coupled to a fiber collimator, which illuminated the detector. The maximum counts (i.e. the peak) were plotted as a function of input power and are shown in Fig. \ref{fig:saturation_characteristics}. From the saturation characteristics, we estimate that the detector starts to deviate from a linear rise at 42000 ADU. We estimated the flux in the brightest pixel by fitting a 2D Gaussian to the beam profile. Thus, the flux where the detector starts to deviate from a linear rise is 1.3 $\mu$W pixel$^{-2}$.

%Where does the detector saturate?
%Gaussian profile, area under the curve = total power from the fiber. Then calculate the flux at the peak pixel. That will give you the ADU to flux conversion and the value of flux (mW/sq. pixel) at which the pixel saturates (or leaves the linear regime of the detector). 

\subsection{Cross-dispersion optics}

\begin{figure} [tbp]
   \begin{center}
   \begin{tabular}{c} %% tabular useful for creating an array of images 
   \includegraphics[height=6cm]{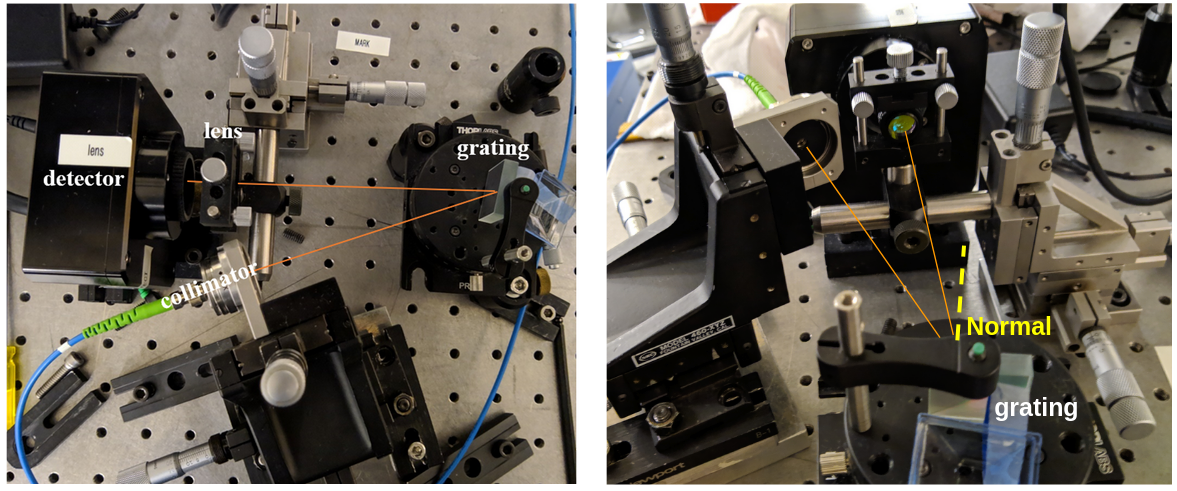}
   \end{tabular}
   \end{center}
   \caption[Grating Disperser] 
%>>>> use \label inside caption to get Fig. number with \ref{}
   { \label{fig:grating_disperser} 
The cross-dispersion setup using the grating, including the collimator, grating, lens, and detector. The
orange line show the path of the light, starting at the collimator and ending at the detector.}
   \end{figure} 

\begin{figure} [tbp]
   \begin{center}
   \begin{tabular}{c} %% tabular useful for creating an array of images 
   \includegraphics[height=5.3cm]{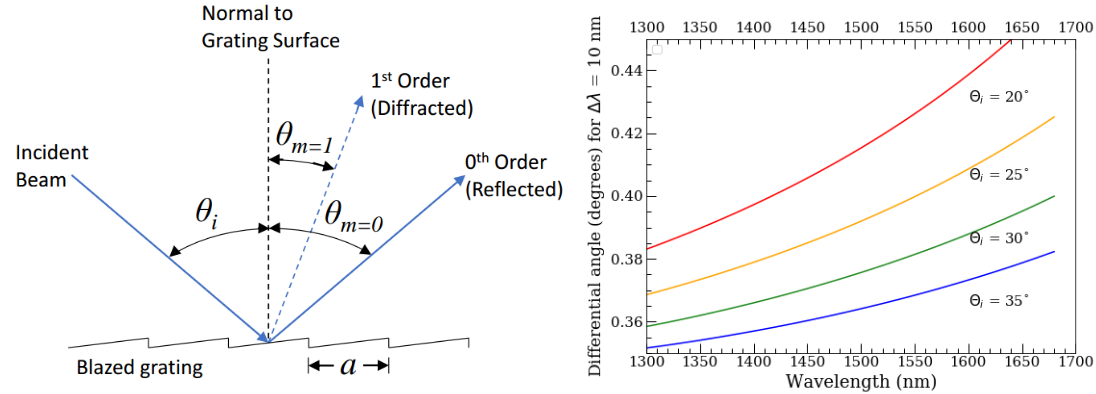}
   \end{tabular}
   \end{center}
   \caption[Blazed grating calculations] 
%>>>> use \label inside caption to get Fig. number with \ref{}
   { \label{fig:blazed_grating} 
Left: A schematic of a blazed grating showing the angles of the incident beam, zeroth order beam, and first
order beam (Reference: Thorlabs.com). Right: The computed variation of differential angle in degrees ($\Delta \theta_{m}$) for $\Delta \lambda$ = 10 nm as a function of wavelength and incidence angle. }
   \end{figure} 

Next, we designed and tested a cross-dispersion setup using a collimated fiber as a source and a blazed grating as a cross-disperser. As described earlier, the AWG disperses the light in the vertical direction and the cross-disperser separate the order in horizontal direction. The setup of the cross-dispersion system in shown in Fig. \ref{fig:grating_disperser}. We use an off-the-shelf NIR grating from Thorlabs \footnote{Datasheet here: https://www.thorlabs.com/thorproduct.cfm?partnumber=GR25-0616}. The blaze angle and the groove density are 28$^{\circ}$41' and 600/mm, respectively. The angle of the diffracted light of spectral order $m$ for a blazed grating is given by the grating equation:

\begin{equation}
    a[sin(\theta_{m}) + sin(\theta_{i})] = m\lambda
\end{equation}

\begin{figure} [tbp]
   \begin{center}
   \begin{tabular}{c} %% tabular useful for creating an array of images 
   \includegraphics[height=7cm]{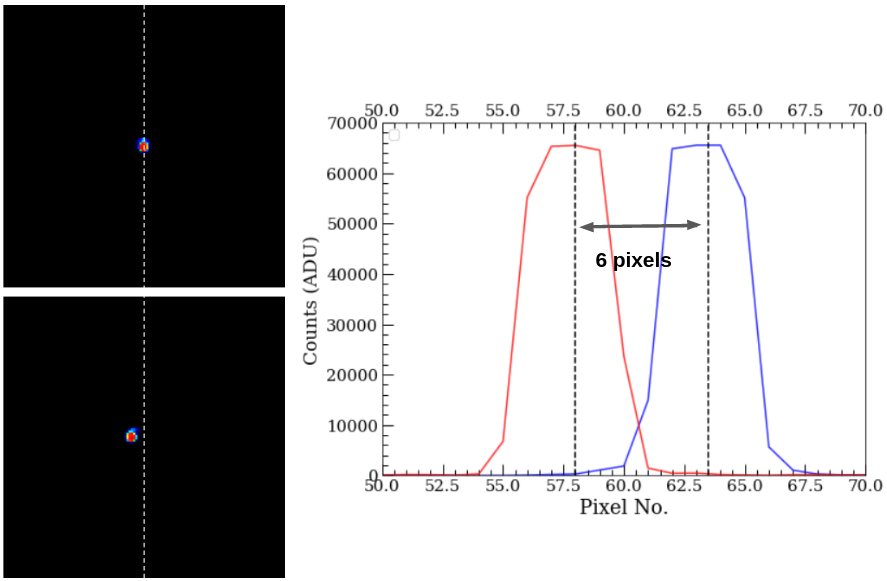}
   \end{tabular}
   \end{center}
   \caption[Simulation results] 
%>>>> use \label inside caption to get Fig. number with \ref{}
   { \label{fig:dispersion_result} 
Left: Two images showing the light output from the grating setup at 1355 nm and 1365 nm. The beam has a diameter of about 5 pixels and has a dispersion of about 6 pixels between the 1355 nm beam  and the 1365 nm beam. Right: The same result with the cross-sections of the beams of 1355 nm (blue) and 1365 nm (red) wavelengths.}
   \end{figure}

\begin{figure} [tbp]
   \begin{center}
   \begin{tabular}{c} %% tabular useful for creating an array of images 
   \includegraphics[height=6cm]{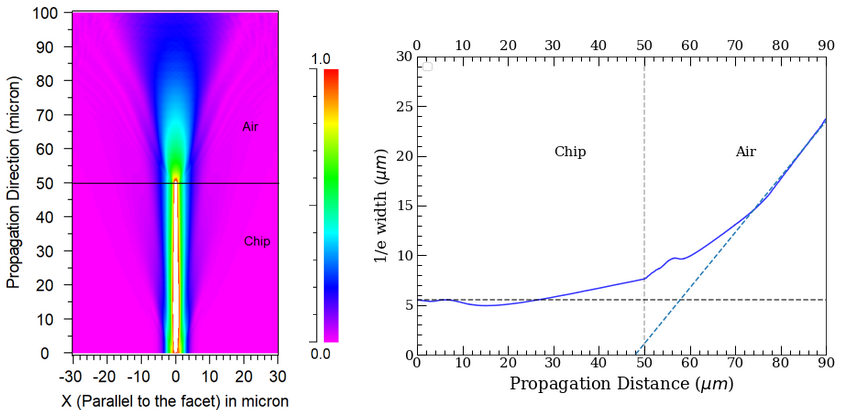}
   \end{tabular}
   \end{center}
   \caption[Simulation of chip to air illumination] 
%>>>> use \label inside caption to get Fig. number with \ref{}
   { \label{fig:chip_to_air} 
Left: Simulated electric field. The transition from chip to air is shown. The waveguide used here is a weakly confined SiN waveguide for the taper end (width = 600 nm , height = 100 nm), similar to the one described in Section \ref{sec:coupling_taper}.   Right: The same result with 1/$e$ width of the field as a function of propagation distance. The half-angle of the emergent beam is computed to be 29.5 $^{\circ}$.}
   \end{figure}   

Here, $a$ is the groove spacing (= 1666 nm), $m$ is the spectral order (= 1), $\theta_{i}$ is the incidence angle and $\theta_{m}$ is the angle of the diffraction ray. Note that $\theta_{i}$ and $\theta_{m}$ are both positive if they are on the same side of the normal. A blazed grating schematic is shown in Fig. \ref{fig:blazed_grating}. For cross-dispersion, the differential angle ($\Delta \theta_{m}$) for $\Delta \lambda$ = 10 nm (which is the span or FSR of one spectral order of the AWG). $R\Delta\theta$ (for $\Delta\theta$ in radians) gives the distance between the two separated orders of the AWG on the detector plane. Here $R$ is the distance between the grating and the detector. Figure \ref{fig:blazed_grating} shows the differential angle ($\Delta \theta_{m}$ in degrees) as a function of wavelength and incidence angle. 

To ensure appropriate mounting and to accommodate maximum number of spectral orders, we select $\theta_{i}$ = 30$^{\circ}$. Using a converging lens of magnification 0.14, we achieve a dispersion of approximately 120 $\mu m$ or 6 pixels and a beam diameter of approximately 4.5 pixels, as shown in Fig. \ref{fig:dispersion_result} for $\lambda_1$ = 1355 nm and $\lambda_2$ = 1365 nm. This will allow us to accommodate about 20 spectral orders from the AWG. 

With a compact 3D-printed mounting setup in the future, the distance between the grating and detector will be smaller, making it possible to accommodate more than 20 spectral orders. The current size of the setup is $\sim$ 30 cm $\times$ 30 cm, but most of the space is taken up by the alignment stages on the optical bench. This can be condensed to a 15 cm $\times$ 15 cm form factor (including the AWG) using a custom 3D-printed mount with precise alignment. Thus, the eventual volume of the instrument will be just 15 $\times$ 15 $\times$ 15 cm$^3$, ideal for deployment at any telescope.

\subsection{Future work: Integration with the AWG}
The next step is to incorporate the AWG as the input source instead of the fiber. The collimated light from the AWG will then be dispersed by the blazed grating in orthogonal direction (thus, separating the spectral orders of the AWG) and imaged on the detector. To collimate the AWG output, the opening angle of the emergent beam from the waveguide has to be determined. We simulated the chip-to-air propagation for an output waveguide using {\fontfamily{qcr}\selectfont Rsoft} software and estimated the opening angle to be 59 degrees (i.e. 2 $\times$ half-angle) as shown in Fig. \ref{fig:chip_to_air}. A schematic of the final assembly of the spectrograph with the AWG and cross-dispersion system is shown in Fig. \ref{fig:AWG_to_detector}. Once the AWG chip is packaged and mounted as per the schematic, the system will be ready for its first on-sky test.

\acknowledgments % equivalent to \section*{ACKNOWLEDGMENTS}       
 
The authors thank the University of Maryland NanoCenter for the fabrication expertise and the astrophotonics group at University of Sydney and Maquarie University for providing valuable advice. The authors acknowledge the financial support for this project from the W. M. Keck Foundation, National Science Foundation, and NASA.   
% References
\bibliography{report} % bibliography data in report.bib

\begin{thebibliography}{10}

\bibitem{bland2009astrophotonics}
Bland-Hawthorn, J. and Kern, P., ``Astrophotonics: a new era for astronomical
  instruments,'' {\em Optics Express}~{\bf 17}(3),  1880--1884 (2009).

\bibitem{bland2017astrophotonics}
Bland-Hawthorn, J. and Leon-Saval, S.~G., ``Astrophotonics: molding the flow of
  light in astronomical instruments,'' {\em Optics express}~{\bf 25}(13),
  15549--15557 (2017).

\bibitem{ellis2017astrophotonics}
Ellis, S., Kuhlmann, S., Kuehn, K., Spinka, H., Underwood, D., Gupta, R.,
  Ocola, L., Liu, P., Wei, G., Stern, N., et~al., ``Astrophotonics: the
  application of photonic technology to astronomy,'' in [{\em Integrated
  Optics: Physics and Simulations III}{\nolinebreak\hspace{0.1em}]},   {\bf
  10242},  102420O, International Society for Optics and Photonics (2017).

\bibitem{gatkine2019astrophotonic}
Gatkine, P., Veilleux, S., and Dagenais, M., ``Astrophotonic spectrographs,''
  {\em Applied Sciences}~{\bf 9}(2),  290 (2019).

\bibitem{Gatkine2019State}
Gatkine, P., Veilleux, S., Mather, J., Betters, C., Bland-Hawthorn, J., Bryant,
  J., Cenko, S.~B., Dagenais, M., Deming, D., Ellis, S., Greenhouse, M.,
  Harris, A., Jovanovic, N., Kuhlmann, S., Kutyrev, A., Leon-Saval, S., Madhav,
  K., Moseley, S., Norris, B., Rauscher, B., Roth, M., and Vogel, S., ``State
  of the profession: Astrophotonics,'' {\em Bulletin of the AAS}~{\bf 51} (9
  2019).
\newblock https://baas.aas.org/pub/2020n7i285.

\bibitem{blind2017spectrographs}
Blind, N., Coarer, E.~L., Kern, P., and Gousset, S., ``Spectrographs for
  astrophotonics,'' {\em arXiv preprint arXiv:1707.01669}  (2017).

\bibitem{bland2006instruments}
Bland-Hawthorn, J. and Horton, A., ``Instruments without optics: an integrated
  photonic spectrograph,'' in [{\em Ground-based and Airborne Instrumentation
  for Astronomy}{\nolinebreak\hspace{0.1em}]},   {\bf 6269},  62690N,
  International Society for Optics and Photonics (2006).

\bibitem{cvetojevic2012first}
Cvetojevic, N., Jovanovic, N., Betters, C., Lawrence, J., Ellis, S., Robertson,
  G., and Bland-Hawthorn, J., ``First starlight spectrum captured using an
  integrated photonic micro-spectrograph,'' {\em Astronomy \&
  Astrophysics}~{\bf 544},  L1 (2012).

\bibitem{harris2012applications}
Harris, R.~J. and Allington-Smith, J., ``Applications of integrated photonic
  spectrographs in astronomy,'' {\em Mon. Not. R. Astron. Soc.}~{\bf 428}(4),
  3139--3150 (2012).

\bibitem{gatkine2017arrayed}
Gatkine, P., Veilleux, S., Hu, Y., Bland-Hawthorn, J., and Dagenais, M.,
  ``Arrayed waveguide grating spectrometers for astronomical applications: new
  results,'' {\em Optics Express}~{\bf 25}(15),  17918--17935 (2017).

\bibitem{zhu2016arbitrary}
Zhu, T., Hu, Y., Gatkine, P., Veilleux, S., Bland-Hawthorn, J., and Dagenais,
  M., ``Arbitrary on-chip optical filter using complex waveguide bragg
  gratings,'' {\em Applied Physics Letters}~{\bf 108}(10),  101104 (2016).

\bibitem{hu2020integrated}
Hu, Y., Xie, S., Zhan, J., Zhang, Y., Veilleux, S., and Dagenais, M.,
  ``Integrated arbitrary filter with spiral gratings: Design and
  characterization,'' {\em Journal of Lightwave Technology}  (2020).

\bibitem{jovanovic2017demonstration}
Jovanovic, N., Cvetojevic, N., Norris, B., Betters, C., Schwab, C., Lozi, J.,
  Guyon, O., Gross, S., Martinache, F., Tuthill, P., et~al., ``Demonstration of
  an efficient, photonic-based astronomical spectrograph on an 8-m telescope,''
  {\em Optics express}~{\bf 25}(15),  17753--17766 (2017).

\bibitem{leon2005multimode}
Leon-Saval, S.~G., Birks, T., Bland-Hawthorn, J., and Englund, M., ``Multimode
  fiber devices with single-mode performance,'' {\em Optics letters}~{\bf
  30}(19),  2545--2547 (2005).

\bibitem{horton2007coupling}
Horton, A.~J. and Bland-Hawthorn, J., ``Coupling light into few-mode optical
  fibres i: The diffraction limit,'' {\em Optics Express}~{\bf 15}(4),
  1443--1453 (2007).

\bibitem{ellis2020first}
Ellis, S., Bland-Hawthorn, J., Lawrence, J., Horton, A., Content, R., Roth, M.,
  Pai, N., Zhelem, R., Case, S., Hernandez, E., et~al., ``First demonstration
  of oh suppression in a high-efficiency near-infrared spectrograph,'' {\em
  Monthly Notices of the Royal Astronomical Society}~{\bf 492}(2),  2796--2806
  (2020).

\bibitem{gatkine2018towards}
Gatkine, P., Veilleux, S., Hu, Y., Bland-Hawthorn, J., and Dagenais, M.,
  ``Towards a multi-input astrophotonic awg spectrograph,'' in [{\em Advances
  in Optical and Mechanical Technologies for Telescopes and Instrumentation
  III}{\nolinebreak\hspace{0.1em}]},   {\bf 10706},  1070656, International
  Society for Optics and Photonics (2018).

\bibitem{zhu2016ultrabroadband}
Zhu, T., Hu, Y., Gatkine, P., Veilleux, S., Bland-Hawthorn, J., and Dagenais,
  M., ``Ultrabroadband high coupling efficiency fiber-to-waveguide coupler
  using {Si$_3$N$_4$/SiO$_2$} waveguides on silicon,'' {\em IEEE Photonics
  Journal}~{\bf 8}(5),  1--12 (2016).

\bibitem{gatkine2016development}
Gatkine, P., Veilleux, S., Hu, Y., Zhu, T., Meng, Y., Bland-Hawthorn, J., and
  Dagenais, M., ``Development of high-resolution arrayed waveguide grating
  spectrometers for astronomical applications: first results,'' in [{\em
  Advances in Optical and Mechanical Technologies for Telescopes and
  Instrumentation II}{\nolinebreak\hspace{0.1em}]},   {\bf 9912},  991271,
  International Society for Optics and Photonics (2016).

\bibitem{hu2018characterization}
Hu, Y.-W., Zhang, Y., Gatkine, P., Bland-Hawthorn, J., Veilleux, S., and
  Dagenais, M., ``Characterization of low loss waveguides using bragg
  gratings,'' {\em IEEE Journal of Selected Topics in Quantum Electronics}~{\bf
  24}(4),  1--8 (2018).

\end{thebibliography}
\bibliographystyle{spiebib} % makes bibtex use spiebib.bst

\end{document}